\documentclass[
 aps,
 pre,
 reprint,
 amsmath,amssymb,
 showpacs,
 floatfix
]{revtex4-1}

\usepackage{graphicx}
\usepackage{dcolumn}
\usepackage{bm}
\usepackage{epsfig}

\begin{document}

\preprint{APS/123-QED}

\title{Increasing the attraction area of the global minimum in the binary optimization problem}
\author{Iakov Karandashev}
\author{Boris Kryzhanovsky}
\affiliation{
 Center of Optical Neural Technologies, Scientific Research Institute for System Analysis, Russian Academy of Sciences\\ Vavilova st. 44-2, Moscow, 119333 Russia
}
\date{\today}

\begin{abstract}
The problem of binary minimization of a quadratic functional in the configuration space is discussed. In order to increase the efficiency of the random-search algorithm it is proposed to change the energy functional by raising to a power the matrix it is based on. We demonstrate that this brings about changes of the energy surface: deep minima displace slightly in the space and become still deeper and their attraction areas grow significantly. Experiments show that this approach results in a considerable displacement of the spectrum of the sought-for minima to the area of greater depth, and the probability of finding the global minimum increases abruptly (by a factor of $10^3$ in the case of the $10 \times 10$ Edwards-Anderson spin glass).
\end{abstract}


\maketitle

\section{Introduction}
Discrete optimization plays a central role in many engineering problems as well as in fundamental science. One major open problem concerns the nature of the energy landscape of optimization problems. It is generally agreed that these energy landscapes are rugged. Finding the ground states of these disordered systems is generally an NP-hard problem so efficient algorithms are particularly called for \cite{Houdayer}. 

The goal of this paper is to make a random-search procedure used in binary-minimization problems more effective. This sort of problem involves minimization of a quadratic functional $E_1 (S)$ built around a particular $N \times N$ matrix $T$ in an $N$-dimensional configuration space of states $S = (s_1 ,\;s_2 \,,\,...,\,s_N )$ where the discrete variables are $s_i  =  \pm 1\;$, $i = \overline {1,N} $.

Minimization in a configuration space is quite different from minimization in a continuous-variable space where it can be solved with the help of conventional methods. Firstly, when we deal with binary variables, the quadratic functional has many extremes even if the matrix $T$ is positive definite, the number of extremes growing exponentially with the dimensionality of the space. Secondly, the problem of binary minimization is NP-complete in the general case, i.e., it does not have a polynomial-time solution algorithm. For this reason heuristic methods are used to solve it. Similarly to the alternating-variable (gradient) descent method, which can be applied to functional minimization in a continuous space, the asynchronous dynamics of the Hopfield neural net \cite{Hopf82,Hopf85} is used in a discrete space. This kind of network keeps decreasing the value of the functional until a local minimum is reached. When we use a random-search algorithm, many randomly chosen initial states are tried to start the neural-net dynamics until a deep enough minimum is reached. Because of the large number of local minima, this method does not work efficiently, though it is an acceptable solution. 

Despite these difficulties, heuristic methods have found wide use in binary-optimization problems. The successful solution of the traveling salesman problem \cite{Hopf85} with the Hopfield neural network introduced neural-net methods in graph theory \cite{Fu}, image processing \cite{Poggio} and other fields \cite{Mulder, Wu, Pinkas, Smith, Joya, Liers, Litin, Boettecher}. The reasons for such a successful application are studied in \cite{Kr2005a} where the probability of finding a minimum in random search is shown to grow exponentially with the depth of the minimum. In particular, when a neural network is initialized at random, it is very likely to converge to a state corresponding to the global minimum or to a local minimum whose depth is comparable to that of the global minimum \cite{Kr2005b, Kr2009}. The probability that the network converges to a relatively shallow local minimum is exponentially small. All this means that the neural network is most likely to find one of the suboptimal solutions (local minimum) if not the optimal solution (global minimum).

Usually, one tries to make the random-search procedure more efficient by changing the algorithm of the descent down the energy landscape described by functional $E_1 (S)$. A good review of these methods is given in \cite{HartmannBook2004, Duch, HartmannBook2001}. In contrast to that approach, we alter the energy landscape itself  in order to increase the radius of the attraction area of the global minimum (and other minima that are almost as deep as the global one). Following the theory developed in \cite{Kar2009}, we have considered the simplest kind of alteration which involves raising the matrix $T$ to the power $k\;$($k = 2,\,3,\,...\;$). The approach proved quite effective: the change of the surface increases the probability of finding the global minimum by $10^3  - 10^4$ times and makes the spectrum of the sought-for minima move far into a deeper-depth region.

We substantiate the efficiency of this algorithm only for the case of matrices whose elements can be considered as random independent variables (matrices of the Edwards-Anderson model, matrices of uniformly or normally distributed elements (Sherrington-Kirkpatrick model), etc.). Application of the algorithm for other types of matrix will be heuristic.

In the following section, we give the general framework in which we work. Section III postulates some necessary theorems concerning the depth and the shape of the local minima. Sections IV and V introduce the double descent algorithm with a matrix to the power $k$ (DDK algorithm) and describe the main features and grounds of the energy landscape transformation. Finally, in Section VI we explain how the algorithm behaves in practice on test problems.

\section{Standard random-search algorithm}
The problem of binary minimization can be formulated as follows: there is an $N \times N$ matrix $T_{ij} $, the goal is to find an $N$-dimensional configuration vector $S_m  = (s_1^{(m)} ,\;s_2^{(m)} \,,\,...,\,s_N^{(m)} )$, $s_i^{(m)}  =  \pm 1\;$, $i = \overline {1,\,\;N} $ that ensures a minimum for the energy functional $E_1 (S)$:	
\begin{eqnarray}
E_1 (S) =  - c_1 \sum\limits_{i = 1}^N {\sum\limits_{j = 1}^N {T_{ij} s_i s_j } },\qquad c_1 = \frac{1}{N^2  \sigma _T}\label{mainfunc}
\end{eqnarray}
where $c_1$ is just a normalization coefficient introduced to allow us to correctly compare the results for different matrices (see Appendix);  $\sigma _T $ is  the standard deviation of the elements of $T$. Without loss of generality, we assume that $T_{ij} $  is a symmetric zero-diagonal matrix ($T_{ij}  = T_{ji} $ , $T_{ii}  = 0$, $\forall i,j = \overline {1,N} $) with zero mean value ($\overline {T_{ij}}=0$).

For minimization we use the Hopfield model \cite{Hopf82} which is the basis of most of today's algorithms of binary optimization. The model is a system of $N$ spins whose interactions are governed by the energy functional $E_1 (S)$. Only conventional (asynchronous) behavior of the Hopfield model is considered: we compute the local field that acts on an arbitrary spin ($i$-th spin):
\begin{equation}
h_i  = \sum\limits_{j \ne i}^N {T_{ij} s_j } .
\label{localfield}
\end{equation}
If the spin is in an unstable state ($h_i s_i  < 0$), the spin changes its state in accordance with the decision rule $s_i  ={\mathop{\rm sgn}} \,h_i $. The procedure is applied successively to all spins until the network comes to a stable state $S_m $. We should point out that in the asynchronous behavior the state of only one spin changes at each moment of time. This is in contrast to synchronous behavior where all spins change their states at the same time. The dynamics under consideration ($h_i  \sim  - \partial E_1 (S)/\partial s_i $) is no other than descending the energy surface $E_1 (S)$ which resembles the alternating-variable (gradient) descent in a continuous space.

A decrease of energy accompanying each overturn of an unstable spin guarantees that the process described by asynchronous dynamics will bring the system to a stable energy-minimum state after a finite number of steps. Of course, this minimum is very likely to be a local one despite our wish to find the deepest (global) minimum of the functional. That is why we have to turn to random search which involves descents from different random initial configurations repeated until the minimum of a specific depth (probably a global minimum) is reached.

\begin{figure}
\includegraphics[width=8.6cm]{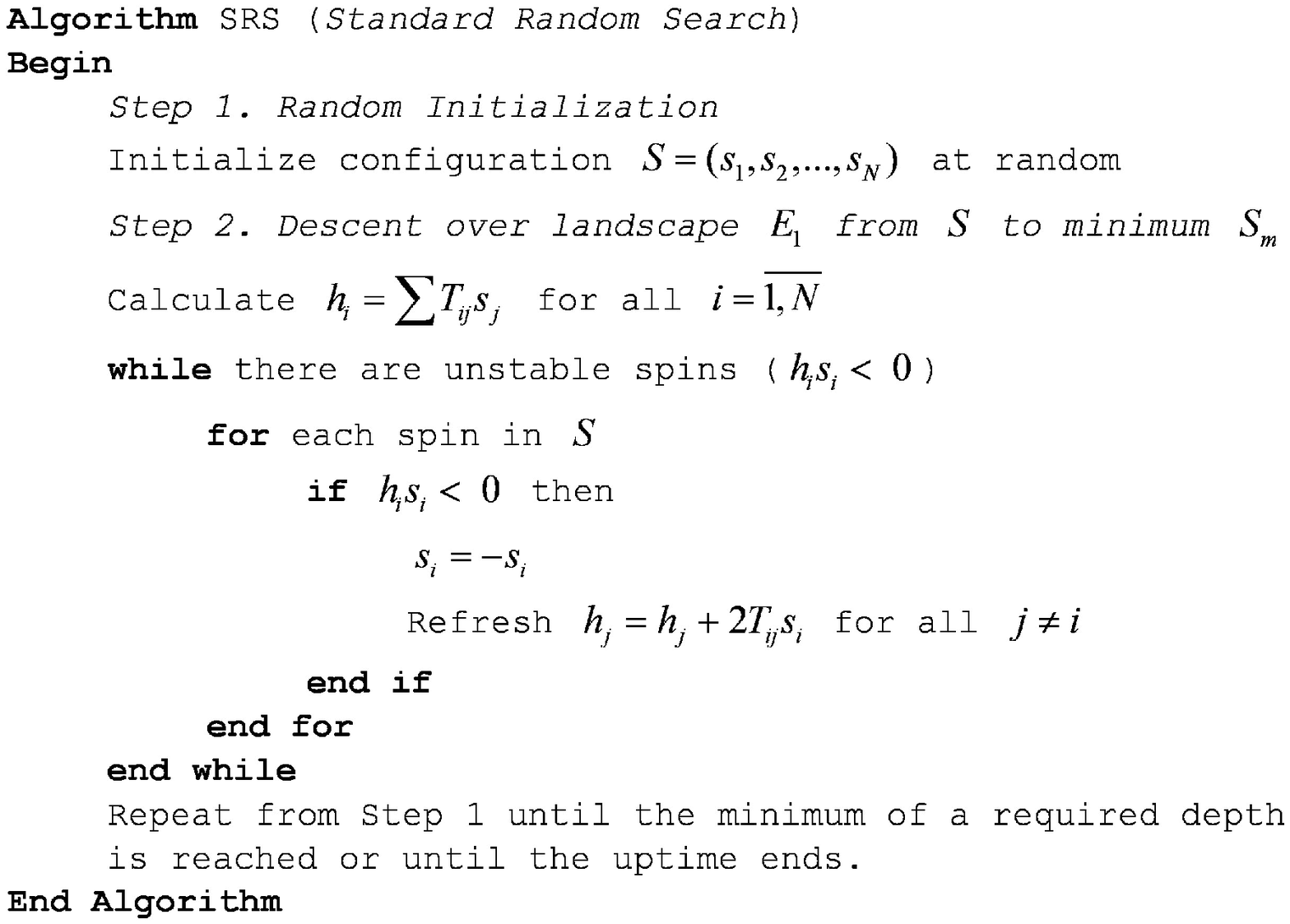}
\caption{\label{figSRS}Conventional random-search algorithm using the asynchronous neural-net dynamics}
\end{figure}

An example of a conventional random-search algorithm is presented in Fig.~\ref{figSRS}. Below we will denote it as SRS (Standard Random Search) and use it for comparison with the algorithm we offer. Though the SRS algorithm is simple, the descent from a single random configuration requires about $O_{SRS}  \approx 2N^2 $ operations for Sherrington-Kirkpatrick model and $O_{SRS}  \approx 8N $ for Edwards-Anderson model with $4N$ nonzero matrix elements $T_{ij}$. Moreover, the number of local minima grows exponentially with $N$ and therefore the probability to find the global minimum falls exponentially. The efficiency of the random-search procedure is usually characterized by three parameters: the probability of finding the global minimum, the search time for a given range of energies, and the average depth of sought-for minima.

\section{Basic relations}
Before coming to the question of surface transformation, let us define the basic relations involving the depth of a global (local) minimum. The first relation deals with the limitation on the depth of a minimum.  Let $S_0  = (s_1^{(0)} ,\;s_2^{(0)} \,,\,...,\,s_N^{(0)} )$ be a configuration corresponding to global minimum $E_0  = E_1 (S_0 )$. Let us single out a term $T_0 $ from $T$. $T_0 $ is responsible for the minimum's appearing. For this purpose let us rewrite $T$ as	
\begin{equation}
T = T_0  + T_1, \label{T}
\end{equation}
where 
\begin{equation}
T_0  = r_0 \sigma _T S_0^ +  S_0, \qquad T_1  = T - T_0. \label{T0} 
\end{equation}
The weight $r_0 $ can be found from the condition that there is no correlation between the elements of $T_0 $ and $T_1 $. By calculating the covariance of these matrix elements and setting it to zero, we find that
\begin{equation}
r_0  =  - E_0. \label{E0}
\end{equation}
This expression establishes the relation between $r_0$ and the depth of the minimum $E_0 $. The variances of the elements of $T_0 $ and $T_1 $ have the form $\sigma _0^2  = r_0^2 \sigma _T^2 $ and $\sigma _1^2  = \sigma _T^2  - \sigma _0^2 $, which verifies that we managed to represent a random matrix $T$ as the sum of two independent matrices $T_0 $ and $T_1 $. In addition, from (\ref{T}) - (\ref{E0}) it follows that $S_0 {\kern 1pt} T_1 \,S_0^ +   = 0$. This relation evidences that the contribution which $T_1 $ makes to $E_0 $ is strictly zero, that is, the minimum at $S_0 $ is solely the contribution of matrix $T_0 $.

Let us follow \cite{Kr2008} and continue expanding (\ref{T}). Let the configuration $S_1 $ be an extreme of a quadratic functional built around $T_1 $. Similarly to (\ref{T}), let us present $T_1 $ as $T_1  = r_1 \sigma _T S_1^ +  S_1  + T_2 $ and define the statistical weight $r_1 $ using zero covariance of the elements of $S_1^ +  S_1 $ and $T_2 $. Repeating this procedure, we can get the matrix in the form of a series of weighted exterior products of random configuration vectors:
\begin{equation}
T = \sigma _T \sum\limits_{m = 0}^\infty  {r_m S_m^ +  S_m } , \qquad \sum {r_m^2 }  = 1 . \label{expansion}
\end{equation}
All the conclusions drawn in \cite{Amit} with the aid of statistical physics hold true for this sort of matrix. In particular, we can state that any of vectors $S_m $ entering the expansion of $T$ will be a minimum of the functional (\ref{mainfunc}) only when its weight $r_m $ is greater than the threshold value
\begin{equation}
r_c  = \frac{1}{{2\sqrt {\alpha _c N} }} , \label{rc}
\end{equation}
where $\alpha _c  \approx 0.138$ is the threshold value of the load parameter \cite{Amit}. First of all, the statement refers to the point $S_0 $ which is the minimum of the functional (1) by definition and for which the following relations are true	
\begin{equation}
1 \ge r_0 \ge r_c ,\qquad  - 1 \le E_0  \le E_c ,\qquad E_c  =  - r_c . \label{restriction}
\end{equation}
The bounding of $r_0 $ above is evident: when $r_0  \to 1$, we have $T = T_0 $, i.e., $T$ is formed as an exterior product of $S_0 $ and $E_1 (S)$ has a single minimum at point $S_0 $. This limit is of no interest because the finding of its global minimum presents no difficulty. In most cases the minimization problem faces the situation when $r_0\sim r_c $ and $E_0\sim E_c$ and the probability to find the global minimum is very low. It is this situation ($r_0 \ll 1$) that we will examine.

The second necessary expression relates the depth and width of a minimum. As shown in \cite{Kr2005a}, the width of a minimum $E_0 $ increases with its depth. Correspondingly, the probability to find this minimum grows exponentially as $P(E_0 )\sim\exp \left( { -NE_c^2 /E_0^2 } \right)$. Such a rigid connection between the depth of the minimum and the probability of finding it can be easily understood from the following consideration. The probability of finding a global minimum in the course of a random search is no other than the ratio between the number of points located within its attraction area and the total number of points ($2^N$) in the $N$-dimensional space. It is possible to show that the shape of the energy surface around the minimum is described \cite{Kr2009} by the expressions
\begin{equation}\label{shape}
\begin{split}
\bar E_0 (n) &\approx E_0 \left( {1 - \frac{{2n}}{N}} \right)^2,\\
\sigma _0 (n) &\approx \frac{2}{N}\sqrt {\frac{{2n}}{N}\left( {1 - \frac{n}{N}} \right)},
\end{split}
\end{equation}
where $\bar E_0 (n)$ and $\sigma _0 (n)$ are the mean value and standard deviation of energy in the $n$-vicinity of the minimum (at points $S_0^{(n)} $ located distance $n$ away from $S_0 $ according to Hamming). Expressions (\ref{shape}) are obtained in the limit $N \to \infty$ from the exact expressions given in Appendix. The energy surface (\ref{shape}) is very smooth because the dispersion of energy in the n-vicinity approaches zero with increasing dimensionality ($\sigma \sim 1/N$). Moreover, differing in depth and width, all minima have the same shape determined by expression (\ref{shape}). As seen from (\ref{shape}) the width of a minimum is proportional to its depth $E_0 $. This means that the number of points within the attraction area grows exponentially with the depth of the minimum.

All of the above conclusions have to do with the configuration $S_0 $ corresponding to the global minimum. However, they are true for any extremum $S_m $ of $E_1 (S)$: $1 \ge r_m  \ge r_c $ and $E_c  \ge E_m  \ge  - 1$ hold for minima, and $r_c  \le - r_m  \le 1$ and $\left| {E_c } \right| \le E_m  \le 1$ hold for maxima.

The result is that we have determined two formulas: a) the deeper the minimum $E_0 $, the greater the weight $r_0 $ of an addition to configuration $S_0 $ in the initial matrix $T$ and the higher the probability of finding the minimum, b) the point $S_0 $ can be a minimum only if $r_0  \ge r_c $, that is,  if the depth of the minimum $\left| {E_0 } \right|$ is greater than the threshold value $\left| {E_c } \right|$. These formulas set the direction of our efforts to improve the efficiency of the random-search algorithm: the energy surface should be transformed in so way that the global minimum will become deeper and, therefore, the probability to find it will grow. This purpose can be reached by raising $T$ to the power $k$ as it is shown in the next section.

\section{Surface transformation}
The surface defined by $E_1 (S)$ can be transformed only by changing the corresponding matrix. Let us put $M = (1 - z)T +zT^k $ into (\ref{mainfunc}). Here, $T^k $ is the matrix resulting from raising $T$ to the power $k$ ($k= 2,3,...$) and zeroing its diagonal elements. Let us pass from $T$ to $M = T^k $ by varying a parameter z from 0 to 1. Correspondingly, the surface determined by $E_1 (S)$ transforms into the surface described by $E_k (S)$:
\begin{equation}
E_k (S) =  - c_k \sum\limits_{i = 1}^N {\sum\limits_{j = 1}^N {M_{ij} s_i s_j } } ,    
c_k  = \frac{1}{{N^2 \sigma _M }} ,
\label{newfunc}
\end{equation}
where $\sigma _M $ is the standard deviation of the elements of $M$. Transformation of the global minimum is taken as the basis for all our considerations. It is clear that the surface transformation will shift the global minimum in the space and will change its depth and attraction area size. We will show below that when the exponent $k$ is relatively small ($2 \le k \le 5$) this transformation will change the minimum depth noticeably. The shift of the minimum is relatively small in case $2 \le k \le 5$.

Correspondingly we offer a two-stage algorithm of minimization. The first stage consists of a descent over the surface $E_k(S)$ and finding a configuration $S_m^{(k)} $ which brings $E_k (S)$ to a minimum. The second stage involves a correction: we descend over the surface $E_1 (S)$ from point $S_m^{(k)} $ to the nearest minimum $S_m $ of $E_1 (S)$. The method of descending the surface $E_k (S)$ is exactly the same as that described above: we compute the local field at the i-th spin:
\begin{equation}
h_i^{(k)}  = \sum\limits_{j \ne i}^N {M_{ij} s_j } ; \label{newlocalfield}
\end{equation}
and when $\,h_i^{(k)} s_i  < 0$, the spin's state changes in accordance with the decision rule $s_i  = {\mathop{\rm sgn}}\,h_i^{(k)} $.

\begin{figure}
\includegraphics[width=8.6cm]{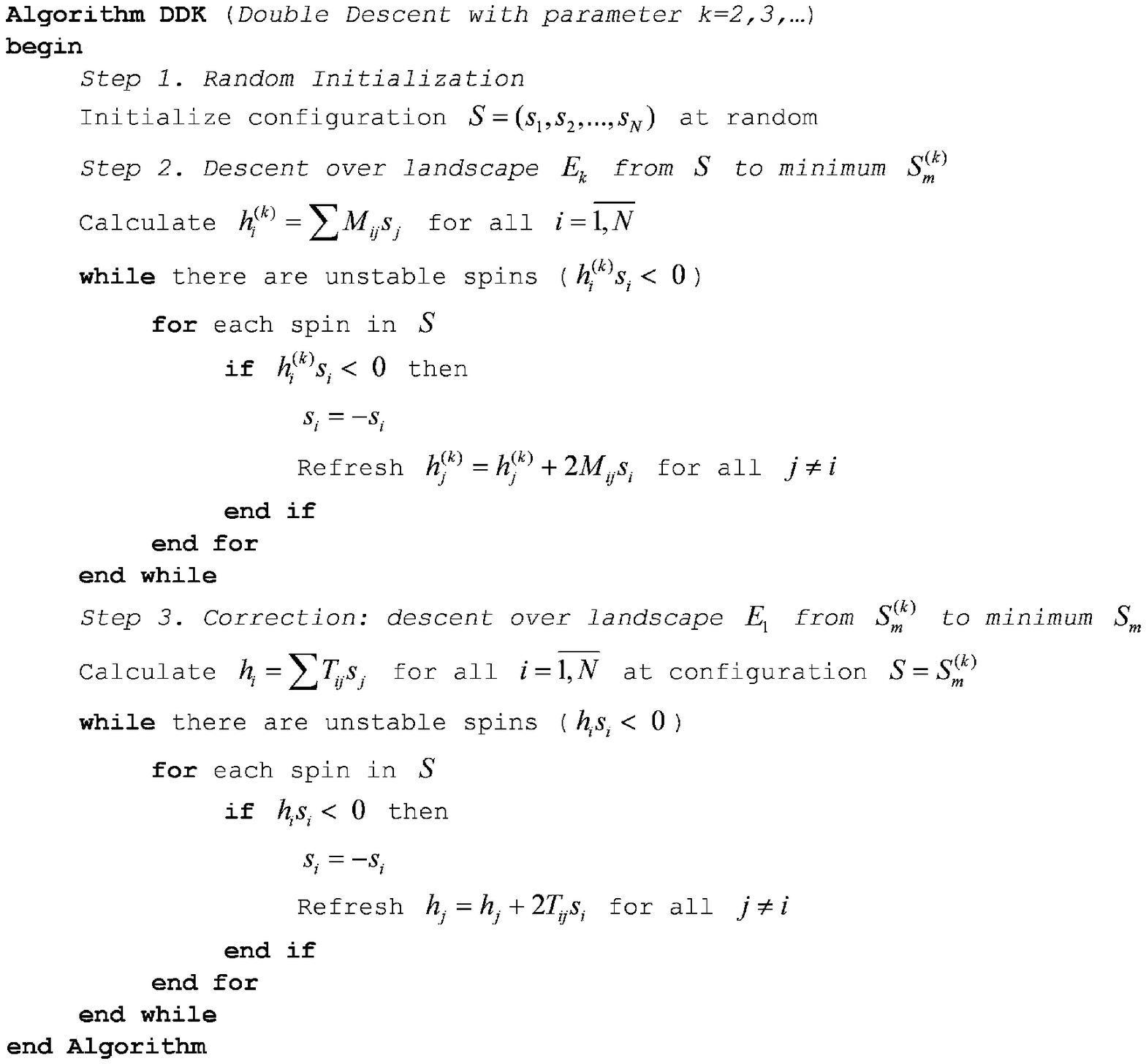}
\caption{\label{figDDK}DDK algorithm of random search using a two-stage descent}
\end{figure}

Using this two-stage descent, we realized the random-search method whose formal algorithm is given in Fig.~\ref{figDDK}. Below we will refer to it as DDK (Double Descent algorithm with parameter $k = 2,3,...$ ) and compare its efficiency with the conventional random-search algorithm SRS. To avoid misunderstanding, note that when $k = 1$, the transformed functional $E_k (S)$ is identical to the original one $E_1 (S)$, and the DDK method does not differ from the common SRS ($DD\,1 \equiv SRS$). 

The number of operations for the DDK algorithm is about twice that of SRS, i.e., $O_{DDK}  \approx 2O_{SRS}$, but we also should remember that we have to spend some time (about $N^3 $ operation) on matrix exponentiation.

\section{The DDK algorithm validation}
To justify the DDK algorithm, we turn to the case when $k = 2$. We will show that the surface transformation makes the global minimum significantly deeper, while its position in binary space changes only slightly.

\subsection{ The deepening of the minima}
Let us first make sure that the transformation of the surface results in the minima becoming deeper. Let us consider the energy $E_{20}  = E_2 (S_0 )$ at the point $S_0 $. Let us make use of (\ref{T}) and represent the matrix $M = T^2 $ as $M = T_0^2  + T_1^2  + (T_0 T_1  + T_1 T_0 )$. Then in view of the relations $S_0 {\kern 1pt} T_1 \,S_0^ +   = 0$ and $\sigma _M  = \sqrt N \sigma _T^2 $ we get from (\ref{newfunc}):
\begin{equation}
E_{20}  =  - \sqrt N E_0^2  + \frac{1}{{N^{5/2} \sigma _T^2 }}\sum\limits_{i = 1}^N {\sum\limits_{j \ne i}^N {(T_1^2 )_{ij} s_i^{(0)} s_j^{(0)} } } . \label{E20}
\end{equation}
It follows from (\ref{E20}) that when $N \gg 1$, $E_{20} $ can be regarded as a normally distributed quantity with mean $\bar E_{20} $ and relatively small standard deviation $\sigma _E $ :
\begin{equation}
\bar E_{20}  =  - \sqrt N E_0^2  ,\qquad  \sigma _E  = (1 - r_0^2 )/N. \label{meanE20}
\end{equation}
Since $\bar E_{20} /E_0  = r_0 \sqrt N  \ge r_c \sqrt N  \approx 1.35$, the minimum should be expected to become deeper after the surface transformation ($E_{20}  < E_0 $). The probability of that event is given by the expression:
\begin{equation}
\Pr \left\{ {E_{20} < E_0 } \right\} = \frac{1}{2}(1 + erf\,\gamma), \label{ProbE20}
\end{equation}
where
\begin{equation}
\gamma  = \frac{{E_0  - \bar E_{20} }}{{\sqrt 2 \sigma _E }} = \frac{{r_0 N(r_0 \sqrt N  - 1)}}{{\sqrt 2 (1 - r_0^2 )}}.
\label{gamma2}
\end{equation}
From $r_0  \ge r_c $ it follows that $\gamma  \ge 0.3\sqrt N $. This means that the probability of the minimum's deepening $\Pr \left\{ {E_{20}  < E_0 } \right\}$ approaches unity with increasing N. In other words, the surface transformation is most likely to cause the minimum to deepen considerably (by a factor of $1.35$ on average), which, according to \cite{Kr2005a}, leads to the probability of finding the global minimum increasing exponentially with $N$.

It should be noted that $\bar E_{20}  =  - \sqrt N E_0^2 $. This allows the conclusion that $E_2 (S)$ reaches its minimum when we deal with both the configuration $S_0 $ which corresponds to the minimum ($E_0  < 0$) of $E_1 (S)$, and the configuration $S_0^* $ which corresponds to the maximum ($E_0  > 0$). Experiments confirm this fact.

\subsection{The shift of the minimum}
Now let us evaluate how far the minimum moves after the surface transformation. The average distance of the shift can be represented as $d_k  = NP$, where
\begin{equation}
P = \Pr \{ s_i^{(0)} h_i^{(k)} < 0\, |\, h_i s_i^{(0)}> 0\} , \label{Pr}
\end{equation} 
is the conditional probability of the spin $s_i^{(0)} $ and the local field $h_i^{(k)} $ having different directions given a configuration of minimum $S_0 $. For our case of $k = 2$ we can get from (\ref{newlocalfield}) the following expression:
\begin{eqnarray}
s_i^{(0)} h_i &=& r_0 \sigma _T  N + \xi_i , \label{sh} \\
s_i^{(0)} h_i^{(2)} &=& r_0^2 \sigma _T^2  N^2 + r_0 N \xi_i + \xi_{2i} , \label{sh2}
\end{eqnarray}
where	
\begin{eqnarray}
\xi_i  &=& \sum\limits_{j \ne i} {(T_1)_{ij} s_i^{(0)} s_j^{(0)} } , \label{xii} \\
\xi_{2i}  &=& \sum\limits_{j \ne i} {(T^2_1)_{ij} s_i^{(0)} s_j^{(0)} } . \label{xi2i}
\end{eqnarray}
are normally distributed quantities with zero means and standard deviations $\sigma _{\xi_i} =  \sigma _{T_1}\sqrt{N/2}$ and $\sigma _{\xi_{2i}} = \sigma _{\xi_i}^2$. This allows us to use the error function to express the probability (\ref{Pr}). It can be shown that
\begin{equation}
P = \frac{1}{P_0\sqrt{\pi}}\int\limits_0^{+\infty} e^{-(x-\gamma)^2}\operatorname{erfc}(\sqrt{2}\gamma x) dx \label{Prfull}, 
\end{equation}
where $P_0=1+\operatorname{erf}(\gamma)$ and $\gamma = \sqrt{N} r_0$.

It is seen that the shift of the minimum $d_2  = N P$ is not large: given $\sqrt {N}r_0 \ge 1.35$, we get that $d_2  \le 0.026N$. It can be shown that in case $k = 3$, the shift is even smaller.

\subsection{Conclusions and their verification}
The following conclusions can be drawn from expressions (\ref{E20})-(\ref{Prfull}). The surface transformation is most likely to make minima deeper and, therefore, the probability of finding them higher. Moreover, the greater the original depth $\left| {E_0 } \right|$, the greater the average increase of the depth $\bar E_{20} /E_0  = \sqrt N \,\left| {E_0 } \right|$. In other words, deep minima become still deeper and the probability to find them becomes yet higher, while shallow minima become more shallow (if not even disappearing) and the probability to discover them become still lower. The above allows us to expect the spectrum of minima found with the given algorithm to move noticeably towards the global minimum, and the probability of finding it to increase considerably.
\begin{figure}
\includegraphics[width=8.6cm]{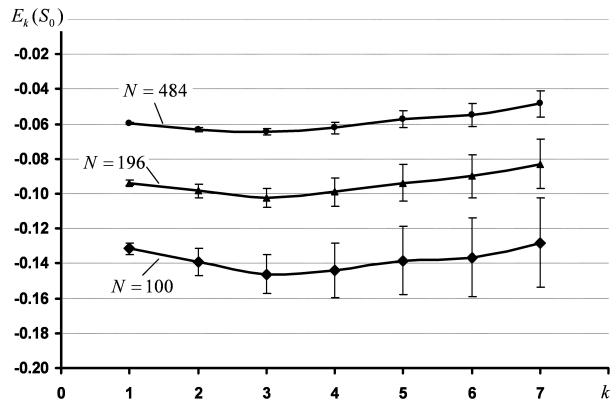}
\caption{\label{figDeepening}The post-transformation deepening and standard deviation of the energy $E_k(S_0)$ when the matrix is raised to a power $k = 2,\,3,...,\,7$. Edwards-Anderson matrices, $N=100, 196, 484$}
\end{figure}
These conclusions have been tested in a few numerical experiments. Fig.~\ref{figDeepening} shows the observed deepening of the global minima. As we see, global minima become deeper for $k=2,\,3,\,4$ and even for $k=5$, but with increasing dispersion. The most average deepening is observed for $k=3$.

\begin{figure}
\includegraphics[width=8.6cm]{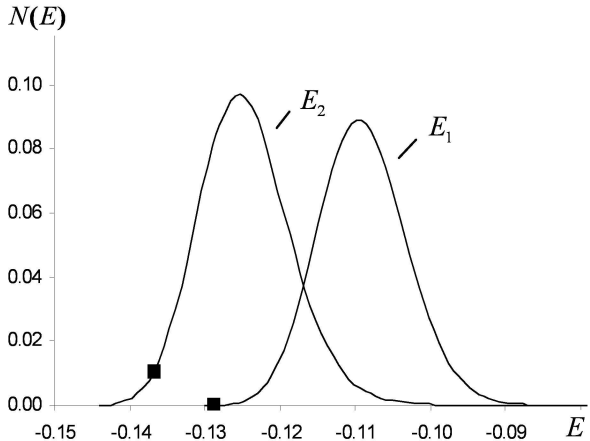}
\caption{\label{figSpectrum}Spectral density of minima of functionals $E_1 (S)$ and $E_2 (S)$. Squares denote energies $E_1 (S_0 )$  and $E_2(S_0 )$ for configuration $S_0 $ which corresponding to the global minimum of functional $E_1 (S)$}
\end{figure}

Since the deepening of minima should lead to an increase in the probability of finding them, the spectrum of the minima of the transformed functional should shift towards greater depths. To confirm this, Fig.~\ref{figSpectrum} shows the spectral densities of the minima of two functional $E_1 (S)$ and $E_2 (S)$ built for $N = 100$. The density $N(E)$ is defined in the following way: $N(E)dE$ is the number of local minima found in the energy interval $[E,\;E + dE]$ and normalized to the total number of minima. We see that the spectrum of $E_2 (S)$ is some distance to the left from that of the original functional $E_1 (S)$, and the minimum at point $S_0 $ has become noticeably deeper. The number of found minima of $E_2 (S)$ is roughly half that of $E_1 (S)$. This shows that we have reached our ends: the transformation of the surface arranged that deep minima become yet deeper and are detected more frequently, while shallow ones are found less frequently or not at all.

\begin{figure}
\includegraphics[width=8.6cm]{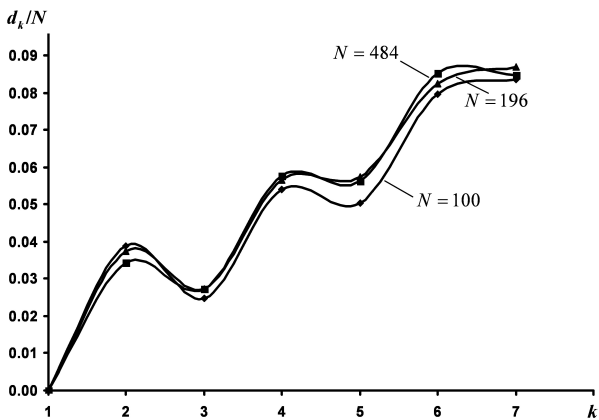}
\caption{\label{figShift}Shift of the global minimum caused by the surface transformation. The curves are experimental data obtained with 2D Edwards-Anderson matrices at $N = 100,\;196,\;484$}
\end{figure}

The shift of the minimum caused by the surface transformation is relatively small. From (\ref{Prfull}) it follows that the smallest shifts are expected for the deepest minima. For example, in case $N=100, k=2$ the experiment shows that for different matrices the shift of global minima varies from 0 to 6 bits and its average value is $d_2 \sim 3.5$ bits, which corresponds well to (\ref{Prfull}). Figure~\ref{figShift} shows the relation between the shift and transformation parameter $k$. It is seen that the smallest shift occurs when $k =3$.

We see that all the theoretical conclusions agree with the experimental results, providing good grounds for the use of the DDK algorithm. In the following sections we will give information on how to apply the algorithm for different types of matrices.

\section{The efficiency of the DDK algorithm}
\subsection{Results for Edwards-Anderson matrices}
The algorithm was tested using mostly matrices of the 2-dimensional Edwards-Anderson model with $N = 10 \times 10$. This type of matrix was chosen for two reasons. Firstly, the standard random search procedure for this sort of matrix is not efficient: when $N = 100$, the probability of finding the global minimum is less than $2 \cdot 10^{ - 6} $. Secondly, in the case of Edwards-Anderson matrices it is possible to use the branch and bound method \cite{Liers} to find the global minimum $S_0 $, which allows us to control the efficiency of the process.

We will determine the efficiency of the DDK algorithm with the help of three parameters: the probability of finding the global minimum, the difference between the average energy of the local minima $\bar E_m $ and the energy $E_0 $ of the global minimum expressed by a dimensionless quantity of discrepancy $\delta E = \left( {\bar E_m  - E_0 } \right)/E_0 $, and the probability of entering the energy interval $[E_0 ,\,\,0.99E_0 ]$.

\begin{figure}
\includegraphics[width=8.6cm]{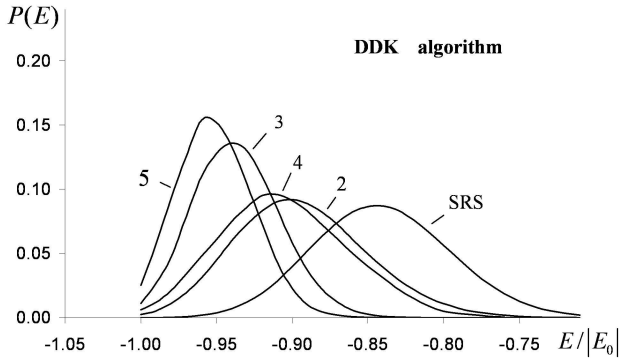}
\caption{\label{figDensity}$P(E)$ is the density of the probability of finding minima of $E_1 (S)$ computed with the aid of the SRS algorithm and DDK algorithm for $k = 2,\;3,\;4,\;5$}
\end{figure}

The DDK algorithm efficiency was evaluated for $k$ from 2 to 7. First of all we watched how the spectrum of local minima of $E_1 (S)$ determined by the DDK algorithm changes when the transformation parameter $k$ grows. The processing of the data allowed us to obtain the probability density $P(E)$ where $P(E)dE$ is the probability of finding minima in the energy interval $[E,\;E + dE]$. Figure~\ref{figDensity} shows the probability density curves determined for a 2D Edwards-Anderson model with $N = 100$, the energy of the global minimum being taken equal to -1. It is seen that the spectrum of minima moves towards the location of the global minimum. Besides, even the probability of finding the global minimum goes noticeably further from zero.

\begin{figure}
\includegraphics[width=8.6cm]{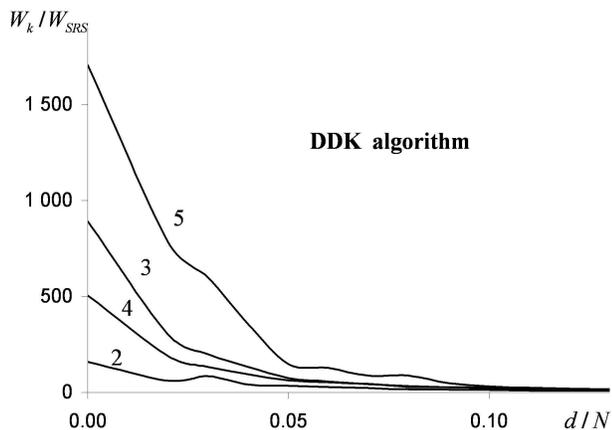}
\caption{\label{figProbDistance}$W_k  = W_k (d)$ is the probability to find minima distance $d$ from the global minimum. $W_k $ is computed with the help of the DDK algorithm for $k = 2,\;3,\;4,\;5$ and divided by $W_{SRS}  = W_{SRS} (d)$ determined using the SRS algorithm}
\end{figure}

In addition, we have investigated how the type of the search algorithm affects the distribution of local minima around the global minimum. The distribution is governed by the probability $W_k  = W_k (d)$ which is computed by the DDK algorithm and determines the probability of finding a local minimum at a point distance $d$ away from the global minimum. The magnitudes of $W_k  = W_k (d)$ do not carry useful information. For this reason Fig.~\ref{figProbDistance} gives $W_k $ normalized to probability $W_{SRS}  = W_{SRS} (d)$ determined by using the SRS algorithm. It is seen that the use of the DDK algorithm increases the probability of finding a local minimum near the global one considerably. In particular, the probability of finding the global minimum ($d = 0$) grows more than three orders of magnitude when $k$ grows.

The experimental results are given in Table~\ref{table1}. The first column holds data for the SRS algorithm. The other columns hold data for the DDK algorithm (for $k = 2 - 5$). To evaluate the efficiency of the algorithm, we picked three characteristics: the first row contains the deviation $\delta E = \left( {\bar E_m  - E_0 } \right)/E_0 $; the second the probability of entering the energy interval $[E_0 ,\,\,0.99E_0 ]$ in close vicinity to the global minimum; the third the probability of hitting the global minimum. It is seen that the results for $k = 5$ little differ from the results for $k = 3$. When $k > 5$, the results become worse. That is why a further increase of $k$ does not make much sense: as $k$ grows, the efficiency of the algorithm falls and computations begin to consume too much time.
\begin{table*}
\caption{\label{table1}Comparison of the DDK algorithm efficiency with the results of using the SRS algorithm performed for a 2D Edwards-Anderson model $N = 10 \times 10$}
\begin{ruledtabular}
\begin{tabular}{p{5cm}ccccc}
 & \textrm{SRS} & \textrm{DD2} & \textrm{DD3} & \textrm{DD4} & \textrm{DD5}\\[8pt]
\textrm{Deviation $\delta E = (\bar E_m - E_0)/E_0$} & $16\%$ & $11\%$ & $6\%$ & $10\%$ & $5\%$\\[8pt]
\textrm{Probability of entering the energy interval $[E_0,\,\,0.99 E_0]$} & $2.8\cdot 10^{-5}$ & $2.3 \cdot 10^{-3}$ & $1.1 \cdot 10^{-2}$ & $5.6 \cdot 10^{-3}$ &  $2.5 \cdot 10^{-2}$\\[8pt]
\textrm{Probability of hitting the global minimum $P_k (d=0)$} & $2.4 \cdot 10^{-6}$ & $3.8 \cdot 10^{-4}$ & $2.1 \cdot 10^{-3}$ & $1.2 \cdot 10^{-3}$ & $4.1 \cdot 10^{-3}$\\
\end{tabular}
\end{ruledtabular}
\end{table*}

\subsection{Results for matrices of Sherrington-Kirkpatrick model}
Unlike sparse Edwards-Anderson matrices, full matrices with evenly (or gaussian) distributed elements allow relatively high chances (about 1\% for matrix dimensionality $N = 100$) to find the global minimum with a conventional neural-net dynamics. However, it was interesting to check how efficiently the method under consideration works with this kind of matrices.

Table~\ref{table2} shows the results of the surface transformation obtained for Sherrington-Kirkpatrick matrices. We dealt with matrices of dimensionality $N = 100$. It is seen that the raising-to-a-power trick also increases the minimization efficiency for this sort of matrix. Indeed, if we use the SRS algorithm, the probability of finding the global minimum is just below one percent, while the raising of the matrix to the third power (DD3 algorithm) allows a tenfold increase of this probability.

\begin{table*}
\caption{\label{table2}The results of using the algorithms SRS and DDK in the case of Sherrington-Kirkpatrick matrices, $N = 100$}
\begin{ruledtabular}
\begin{tabular}{p{5cm}ccccc}
 & \textrm{SRS} & \textrm{DD2} & \textrm{DD3} & \textrm{DD4} & \textrm{DD5}\\[8pt]
\textrm{Deviation $\delta E = (\bar E_m - E_0)/E_0$} & $10\%$ & $5.4\%$ & $2.7\%$ & $4.8\%$ & $1.9\%$\\[8pt]
\textrm{Probability of entering the energy interval $[E_0,\,\,0.99 E_0]$} & $2.7\cdot 10^{-2}$ & $9.3 \cdot 10^{-2}$ & $0.24$ & $0.15$ &  $0.22$\\[8pt]
\textrm{Probability of hitting the global minimum $P_k (d=0)$} & $0.01$ & $0.04$ & $0.12$ & $0.07$ & $0.08$\\
\end{tabular}
\end{ruledtabular}
\end{table*}

\subsection{Improvement of the dynamics. Houdayer-Martin algorithm}
It follows from the above that the increased efficiency of the algorithm is caused by a static change - the alteration of the surface being descended. However, it is possible to modify the dynamics of descent. We found in experiments that the descent often stops in shallow pits just a few bits off the global minimum. In order to avoid this unfavorable effect, we changed the algorithm of descent.

We borrowed a local optimization algorithm from \cite{Houdayer}. This algorithm was inspired from the Kernighan-Lin \cite{Kernighan} algorithm and in \cite{Houdayer} it was adapted to our problem of quadratic functional minimization in binary space. This local search algorithm is similar to Hopfield dynamics. However it is more efficient due to its capability of going out of small cavities in the energy landscape.

Here we use this algorithm in the form proposed in \cite{Houdayer}.
The number of operations that occur during the execution of one start of the Houdayer-Martin dynamics is about $O_{HM}  \approx 10 O_{SRS} $.

The use of this new Houdayer-Martin dynamics of descent instead of Hopfield neural net dynamics improved the efficiency of the DDK algorithm by an order of magnitude. The improved characteristics are shown in Table~\ref{table3} (the SRS and DDK algorithms using a modified Houdayer-Martin dynamics are designated as SRS-HM and DDK-HM here).

\begin{table*}
\caption{\label{table3}Efficiency of modified SRS and DDK algorithms with Houdayer-Martin dynamics of descent. Data correspond to a 2D Edwards-Anderson model with $N = 100, 196$}
\begin{ruledtabular}
\begin{tabular}{p{5cm}ccccc}
 & \textrm{SRS-HM} & \textrm{DD2-HM} & \textrm{DD3-HM} & \textrm{DD4-HM} & \textrm{DD5-HM}\\[8pt]
\hline
\multicolumn{2}{c}{$N=100$} &&&&\\ [8pt]
\textrm{Deviation $\delta E = (\bar E_m - E_0)/E_0$} & $4.7\%$ & $4.0\%$ & $2.6\%$ & $4.0\%$ & $2.5\%$\\[8pt]
\textrm{Probability of entering the energy interval $[E_0,\,\,0.99 E_0]$} & $5.0\%$ & $8.2\%$ & $16.2\%$ & $9.0\%$ &  $17.4\%$\\[8pt]
\textrm{Probability of hitting the global minimum $P_k (d=0)$} & $1.4\%$ & $2.8\%$ & $6.0\%$ & $3.9\%$ & $5.9\%$\\[8pt]\hline
\multicolumn{2}{c}{$N=196$} &&&&\\ [8pt]
\textrm{Deviation $\delta E = (\bar E_m - E_0)/E_0$} & $5.9\%$ & $4.8\%$ & $3.4\%$ & $4.7\%$ & $3.2\%$\\[8pt]
\textrm{Probability of entering the energy interval $[E_0,\,\,0.99 E_0]$} & $0.2\%$ & $0.5\%$ & $2.1\%$ & $0.6\%$ &  $2.0\%$\\[8pt]
\textrm{Probability of hitting the global minimum $P_k (d=0)$} & $4.7\cdot 10^{-5}$ & $2.2\cdot 10^{-4}$ & $4.3\cdot 10^{-4}$ & $3.6\cdot 10^{-4}$ & $6.7\cdot 10^{-4}$
\end{tabular}
\end{ruledtabular}
\end{table*}

\begin{figure}
\includegraphics[width=8.6cm]{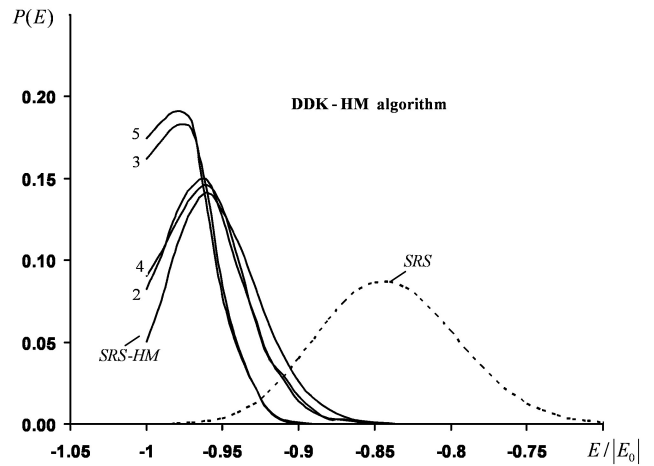}
\caption{\label{figDensityKL}Probability density $P(E)$ found with the help of the modified search algorithms SRS-HM and DDK-HM for $k = 2,\;3,\;4,\;5$. The dashed line corresponds to the probability density computed by the SRS method}
\end{figure}

Figure~\ref{figDensityKL} gives the probability density curves $P(E)$ when we use the modified search algorithms DDK-HM and SRS-HM. Comparing them with the curves in Fig.~\ref{figDensity}, we can see that this modification of the descent dynamics brings about a significant shift of the spectral curves towards the global minimum. It is seen that due to this modification the probability of arriving at a global minimum increased by another order of magnitude. The DD3-HM algorithm gives the best result: the probability to find the global minimum is $2.5 \cdot 10^4$ times higher than for the SRS method.

For Sherrington-Kirkpatrick matrices the gain of using the DDK-HM algorithm instead of the SRS-HM algorithm becomes evident with increasing dimensionality $N$.

\section{Discussion}
Let us again formulate the DDK algorithm. The preparatory stage involves three steps: 1) the original matrix $T$ is made symmetric (if it is not so initially) and the diagonal elements are set to zero; 2) the matrix is raised to the $k$th   power and the diagonal elements of the resulting matrix $M = T^k $ are set to zero; 3) $E_k (S)$ is built using $M$ and formula (\ref{newfunc}). The preparatory stage is followed by the random search procedure which employs the two-stage descent algorithm: a) in the first stage we descend the surface $E_k (S)$ from the randomly chosen original configuration to the nearest local minimum $S_m^{(k)} $ of $E_k (S)$; b) in the second stage we make a correction by going down the surface $E_1 (S)$ from $S_m^{(k)} $ to the nearest local minimum $S_m $ of $E_1 (S)$ which is usually located near point $S_m^{(k)} $. In both stages we can use any dynamics of descent (i.e., any local optimization procedure), e.g., Hopfield's or Houdayer-Martin's, discussed earlier.

Comparison shows that the surface transformation increases the optimization algorithm efficiency significantly. In particular, for Edwards-Anderson matrices of size $N = 196$, the use of the DD3-HM method allows the following improvements: the probability of hitting the global minimum increases by order of magnitude in comparison with the SRS-HM algorithm; the difference between the average energy $\bar E_m $ and the energy of the global minimum decreases by half; and the probability of getting the energy interval around the global minimum $E_m  \in [E_0 ,\,\,0.99E_0 ]$ grows by order of magnitude.

The analysis shows that the power $k$ have an optimal value $k=3$, because the shift of global minimum is minimal (see Fig.~\ref{figShift}) and the deepening is maximal (Fig.~\ref{figDeepening}) at $k=3$. Wnen $k>3$ the dispersion of deepening increases and algorithm becomes unstable: when experimenting with DD5, we could not detect the global minimum for $60\%$ of matrices. That is why one should keep in mind that though Table~\ref{table1} tells us about the higher efficiency of the DD5 algorithm than DD3, these are averaged data and the actual situation may be quite different for a particular kind of matrix.

Adding the Houdayer-Martin algorithm instead of the neural-net dynamics also improves the algorithm considerably. In particular, the DD3-HM algorithm allows hitting the global minimum with an average probability of about $6\% $, which is 30 times higher than the probability ($ \sim 0.2\% $) provided by the DD3 algorithm and 25,000 times greater than the probability ($ \sim 2.4 \cdot 10^{ - 6} $) permitted by the SRS algorithm. Introduction of a Houdayer-Martin approach increases the time of a single descent tenfold (see Table~\ref{table4}). However, this dynamics adds stability to the algorithm: there were not such matrices for which the DD3-HM algorithm failed to find the global minimum.

The above allows the following conclusion: in all experiments the double-descent algorithm (DDK or DDK-HM) brings the system to a configuration that is very close to the global minimum in energy and distance with a very high probability, all characteristics of the algorithm being much better than those of the SRS algorithm. Besides, the superiority of the DDK (DDK-HM) algorithm over the SRS algorithm only grows with increasing dimensionality $N$. The examination of the experimental data shows that the DD3-HM algorithm is the best for practical use: it exhibits the highest probability of finding the global minimum, the greatest stability, and the smallest search time.

\begin{table}
\caption{\label{table4}Comparing the DDK methods in time consumption.  $N = 100$}
\begin{ruledtabular}
\begin{tabular}{lc}
\textrm{Method} & \textrm{The average time per 1000 starts (s)}\\
\colrule
\textrm{SRS} & $0.090$\\
\textrm{DD2} & $0.175$\\
\textrm{DD3} & $0.167$\\
\textrm{DD5} & $0.170$\\[4pt]
\textrm{SRS-HM} & $0.950$\\
\textrm{DD2-HM} & $1.571$\\
\textrm{DD3-HM} & $1.211$\\
\textrm{DD5-HM} & $1.152$\\
\end{tabular}
\end{ruledtabular}
\end{table}

\begin{acknowledgments}
This research was supported by the Russian Foundation for Basic Research (grant 09-07-00159)
\end{acknowledgments}

\appendix*
\section{The shape of the energy landscape}
Consider an arbitrary point $S_0  = (s_1^{(0)} ,\,s_2^{(0)} ,\,...,\,s_N^{(0)} )$ in $N$-dimensional configurational space. The set of all points $\Omega  = \left\{ {S_0^{(n)} } \right\}$ such that $S_0^{(n)} $ is located $n$ bits away from $S_0 $ is called the n-vicinity of $S_0 $. Let us show that as $N \to \infty $ the dispersion of energy in the n-vicinity is negligible and the energy surface around $S_0 $ is determined by the mean value of the energy in the $n$-vicinity.

The components of a vector $S_0^{(n)} $ that belongs to the $n$-vicinity and the components of $S_0 $ are connected by the relation: 
$$s_i^{(n)}  = a_i^{(n)} s_i^{(0)} 
\forall i = \overline {1,N} ,
$$
where ${\bf{a}}_n $ is a vector such that some $N - n$ of its components equal $ + 1$ and the other $n$ components equal $ - 1$.  Using this notation, the energy at the point $S_0^{(n)} $ is:
\begin{equation}
E_0 (n) =  - c\sum\limits_i {\sum\limits_j {T_{ij} s_i^{(0)} s_j^{(0)} } } a_i^{(n)} a_j^{(n)} \chi _{ij} ,
\label{En}
\end{equation}
where $\chi _{ij} $ is the unit antisymmetric tensor. Further, the sum of the energies and the sum of squared energies over the set of all points of the n-vicinity is described by the expressions:
\begin{eqnarray}
\sum\limits_\Omega  {E_0 (n)}  &=&  - c\sum\limits_{i = 1}^N {\sum\limits_{j = 1}^N {K_{ij} T_{ij} s_i^{(0)} s_j^{(0)} } } \label{sumEn}
\\
\sum\limits_\Omega  {E_0^2 (n)}  &=&  - c\sum\limits_{i,j,k,l}^N {K_{ijkl} T_{ik} T_{jl} s_i^{(0)} s_k^{(0)} s_j^{(0)} s_l^{(0)} }  ,
\label{sumsqEn}
\end{eqnarray}
where	
\begin{equation}
K_{ij}  = \sum\limits_\Omega  {a_i a_j \chi _{ij} }  ,
K_{ijkl}  = \sum\limits_\Omega  {a_i a_j a_k a_l \chi _{ik} } \chi _{jl} .
\label{Kijkl}
\end{equation}

Summing (\ref{Kijkl}) over the sphere $\Omega  = \left\{ {S_n } \right\}$ and substituting the result into (\ref{sumEn}) and (\ref{sumsqEn}), we obtain the mean value and the dispersion of energy in the $n$-vicinity:
\begin{equation}
\bar E_0 (n) = E_0 \frac{{(N - 2n)^2  - N}}{{N(N - 1)}} ,
\label{meanEn}
\end{equation}
\begin{equation}
\sigma _0^2 (n) = \frac{{8n(N - n)}}{{N(N - 1)(N - 2)(N - 3)}}\left[ {A\left( {1 - E_0^2 } \right) + B\sigma _{hs}^2 } \right] ,
\label{sigmaEn}
\end{equation}
where $E_0  =  - \sum {h_i s_i^{(0)} } $ is the energy of the system at $S_0 $,  $h_i $ is the local field acting on the $i$-th spin of the configuration $S_0 $, $\sigma _{hs} $ is the standard deviation of $h_i s_i^{(0)} $, and we use the following notation:
\begin{equation}
A = \frac{{4(n - 1)(N - n - 1)}}{{N(N - 1)}} ,
\label{A}
\end{equation}
\begin{equation}
B = 2N\left[ {(N - 2n)^2  - (N - 2)} \right] .
\label{B}
\end{equation}
According to (\ref{meanEn}) the mean value of the energy in the $n$-vicinity behaves parabolically. At $n = 0$ the mean energy equals $E_0 $; with increasing $n$ the mean energy $\bar E_0 (n)$ rises as if the configuration $S_0 $ were surrounded by a funnel with parabolic walls. At $n = {\textstyle{\frac{1}{2}}}(N - \sqrt N )$ the quantity  $\bar E_0 (n)$ vanishes and at $n = N/2$ the parabola reaches its peak. The further trend of the curve is completely symmetric: with increasing $n$ the mean energy goes down parabolically up to energy $E_0 $ at $n = N$ when the configuration turns into mirroring $S_0^*  =  - S_0 $.  

Expressions (\ref{meanEn})-(\ref{sigmaEn}) are true for any point of the configuration space. However, $\bar E_0 (n)$ describes more precisely the shape of the energy surface around the minimum: it follows from (\ref{sigmaEn}) that the energy dispersion is minimal when $S_0 $ is the global minimum because $\sigma _{hs} $ is minimal and $E_0^2 $ is maximal. The energy surface described by $\bar E_0 (n)$ is very smooth because its roughness vanishes with increasing N ($\sigma _0 (n)\sim N^{ - 1} $).

It is not hard to calculate $\sigma _{hs} $ for any point of the space. But when calculating $\sigma _{hs} $ for a minimum $S_0 $ we should take into account that $h_i s_i^{(0)}  > 0$, $\forall i = \overline {1,N} $. Under this condition, we have $2N^3 \sigma _{hs}^2 \sim1$ (the dispersion of $h_i s_i^{(0)} $ at a minimum is half that at an arbitrary point). Using this relation and keeping only the first terms in (\ref{meanEn})-(\ref{sigmaEn}) that are nondecreasing in the small parameter $\sim O(N^{ - 1} )$, we get the approximate expression (\ref{shape}).

Summing (\ref{sumEn})-(\ref{sumsqEn}) over all n-vicinities ($n = \overline {0,N} $), i.e., over the whole space, we have a trivial result: the average value of energy is zero ($\overline {E_0 (n)}  = 0$) and the average square of energy is determined only by the average square of the matrix elements ($\overline {E_0^2 (n)}  = 2c^2 N(N - 1)\overline {T_{ij}^2 } $), which coincides with the dispersion of matrix elements under the condition of zero mean value $\overline {T_{ij} } $.  This explains the meaning of the normalization constant $c$ in the energy functional: due to this normalization the energy value is independent of the sort of the matrix; so we can compare results for different matrices of the same type and also different sorts of matrices.

\bibliography{mybib}

\end{document}